\documentclass[aps,prl,twocolumn,letterpaper,10pt,superscriptaddress]{revtex4-2}

\usepackage{amssymb,amsthm,amsmath,amsfonts,mathcmd,mathrsfs}
\usepackage{graphicx,ulem,enumerate,bm,bbm,mathptmx}
\usepackage[pdftex,dvipsnames,usenames]{xcolor}
\usepackage[colorlinks=true,urlcolor=blue,citecolor=blue,linkcolor=blue]{hyperref}

\usepackage{braket,tensor,tikz}
\usepackage{stmaryrd}
\usepackage{esint}

\newcommand{\me}{\mathrm{e}}

\begin{document}

\title{
    Separability Lindblad equation for dynamical open-system entanglement
}

\author{Julien Pinske}
    \affiliation{Niels Bohr Institute, University of Copenhagen, Blegdamsvej 17, DK-2100 Copenhagen, Denmark}

\author{Laura Ares}
    \affiliation{Institute for Photonic Quantum Systems (PhoQS), Paderborn University, Warburger Stra\ss{}e 100, 33098 Paderborn, Germany}

\author{Benjamin Hinrichs}
    \affiliation{Institute for Photonic Quantum Systems (PhoQS), Paderborn University, Warburger Stra\ss{}e 100, 33098 Paderborn, Germany}

\author{Martin Kolb}
    \affiliation{Institute for Photonic Quantum Systems (PhoQS), Paderborn University, Warburger Stra\ss{}e 100, 33098 Paderborn, Germany}

\author{Jan Sperling}
    \affiliation{Institute for Photonic Quantum Systems (PhoQS), Paderborn University, Warburger Stra\ss{}e 100, 33098 Paderborn, Germany}

\date{\today}

\begin{abstract}
    Providing entanglement for the design of quantum technologies in the presence of noise constitutes today's main challenge in quantum information science.
    A framework is required that assesses the build-up of entanglement in realistic settings.
    In this work, we put forth a new class of nonlinear quantum master equations in Lindblad form that unambiguously identify dynamical entanglement in open quantum systems via deviations from a separable evolution.
    This separability Lindblad equation restricts quantum trajectories to classically correlated states only.
    Unlike many conventional approaches, here the entangling capabilities of a process are not characterized by input-output relations, but separability is imposed at each instant of time.
    We solve these equations for crucial examples, thereby quantifying the dynamical impact of entanglement in non-equilibrium scenarios.
    Our results allow to benchmark the engineering of entangled states through dissipation.
    The separability Lindblad equation provides a unique path to characterizing quantum correlations caused by arbitrary system-bath interactions, specifically tailored for the noisy intermediate-scale quantum era.
\end{abstract}

\maketitle

\paragraph*{Introduction.---}
    
    Quantum entanglement is a key property of composite quantum systems, describing correlations between two or more subsystems that cannot be attributed to any classical, joint description \cite{W89,HH09}.
    Entanglement thus refers to a lack of separation between subsystems and is considered to be the quintessential ingredient for quantum information processing tasks, including quantum teleportation \cite{BB93,R17}, quantum-computational speedup \cite{RB01,RB02}, quantum communication \cite{BS99,B02}, and quantum-error correction \cite{G96,HD07}.

    As a consequence, a primary concern in quantum information science is to verify the actual presence of entanglement, posing an NP-hard problem \cite{I07}.
    Nevertheless, several inseparability criteria have been devised \cite{HH96,T00}, with the construction of entanglement witnesses being one of the most successful approaches in practice \cite{GT09,LK00,SV13,SR17}.
    However, such witnesses are limited to stationary scenarios, and verifying whether a dynamical process itself creates entanglement is a significantly less explored subject \cite{WB06,I09,SW17,AR20}.

    The most general change a state $\rho_0$ can undergo is given by a dynamical map $\rho(t)=\Lambda_t(\rho_0)$, which encompasses processes such as state preparation, noisy propagation, measurements, etc. \cite{K71,BL16}.
    Many dynamical maps are solutions of a quantum master equation in Lindblad form \cite{L76,NR20,BS22},
    \begin{equation}
        \label{eq:Lindblad}
        \frac{d\rho}{dt}=i\left[\rho,H\right]
        +\sum_{m}\Big(L^{m}\rho L^{m\dag}-\frac{1}{2}\left\{L^{m\dag} L^{m},\rho\right\}\Big),
	\end{equation}
    where $[\,\cdot\,,\,\cdot\,]$ and $\{\,\cdot\,,\,\cdot\,\}$ denote the commutator and anti-commutator, respectively, and $\hbar=1$.
    The Hamiltonian $H$ of the system yields a unitary contribution to the overall evolution, and the Lindblad operators $L^{m}$ account for dissipation.
    Once a solution to Eq. \eqref{eq:Lindblad} is obtained for a separable initial state, the question of whether $\rho(t)$ became entangled at any time $t$ is closely related to asking if the dynamical map $\Lambda_t(\rho_0)=\sum_m K^{m}\rho_0 K^{m\dag}$ is separable \cite{VP97,GG12,CV20}. 
    For example, for two qubits $A$ and $B$, the map $\Lambda_t$ is separable if its corresponding Kraus operators $K^m=K^m_A\otimes K^m_B$ can be given as products.

    In this Letter, we derive a novel type of nonlinear Lindblad master equation, enforcing separable dynamics, dubbed \textit{separability Lindblad equation}.
    Comparing its solutions with the possibly entangling dynamics \eqref{eq:Lindblad} allows one to determine the impact of entanglement over the duration of a noisy process.
    This applies even if the initial and final states are not entangled.
    This situation is common in quantum computation, where an initially separable register of qubits is transformed by noisy entangling gates until the final measurement returns a computational state \cite{D00}.
    Our equation yields a separable evolution by constraining the propagation to the tangent space of separable states.
    Unlike other treatments, we do not employ input-output relations but impose separability at every instant of the evolution.
    Our approach is useful in benchmarking protocols for quantum state engineering through dissipation \cite{VWC09}, and we demonstrate this explicitly for two-qubit systems.

\paragraph*{Separable evolution in open quantum systems.---}

    For a sufficiently short time step $\tau$, the evolved state can be written as $\rho(t+\tau)\approx \rho(t)+\tau\frac{d\rho(t)}{dt}$.
    Then, the dynamics governed by Eq. \eqref{eq:Lindblad} can be expressed via a Kraus map
	\begin{equation}
        \label{eq:FormalSolution}
          \rho(t+\tau)=\sum_{m} K^{m}\rho(t)K^{m\dag},
	\end{equation}
	with $K^0=\mathbbm{1} -i\tau\hat{H}$, corresponding to a propagation with a non-Hermitian Hamiltonian $\hat{H}=H-\frac{i}{2}\sum_m L^{m\dag}L^m$, and $K^{m}=\sqrt{\tau}L^{m}$ being associated with quantum jumps.
    Starting with a separable two-qubit state $|\psi_A\psi_B\rangle=|\psi_A\rangle\otimes|\psi_B\rangle$, the state after a time $\tau$ corresponds to an ensemble $\{K^m\ket{\psi_A\psi_B}\}_m$ of pure states that are not necessarily product states anymore.

\begin{figure}
    \includegraphics[width=0.9\columnwidth]{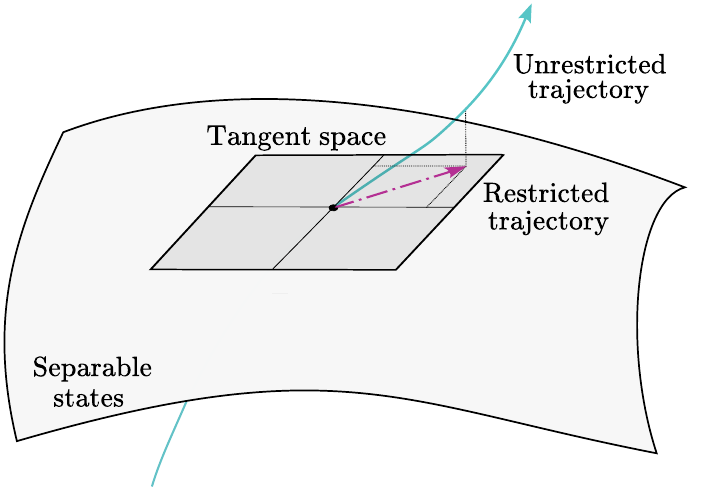}
    \caption{%
        Schematic representation of the tangential approximation.
        In general, the unrestricted evolution takes a state outside the manifold of separable states.
        In the restricted evolution, each infinitesimal time step is projected onto the tangent space of separable states, ensuring the dynamics to contain classical correlations only.
    }\label{fig:tangent}
\end{figure}

    To gauge the impact of entanglement during an open-system evolution, a method is required that yields separable dynamics by limiting the state's dynamics to the separable domain, Fig. \ref{fig:tangent}.
    Expanding $|\psi_A\rangle=\alpha_0|0\rangle+\alpha_1|1\rangle$ and $|\psi_B\rangle=\beta_0|0\rangle+\beta_1|1\rangle$ in terms of orthonormal states $\ket{0}$ and $\ket{1}$, we can introduce a vector tangent to $\ket{\psi}=\ket{\psi_A\psi_B}$ as
	\begin{equation}
        \label{eq:tangentialTP}
    \begin{split}
            |\psi^{\prime}\rangle&=\ket{\psi}+\sum_{a=0,1}\varepsilon_A^a\frac{\partial|\psi\rangle}{\partial \alpha_a}+\sum_{b=0,1}\varepsilon_B^b\frac{\partial|\psi\rangle}{\partial \beta_b},\\
            &=|\psi_{A}\psi_{B}\rangle
        +|\varepsilon_A\psi_B\rangle
        +|\psi_A\varepsilon_B\rangle,\\
    \end{split}
	\end{equation}
    with coefficients $\varepsilon^c_k$ obeying $|\varepsilon^c_k|<\varepsilon$, and vectors $|\varepsilon_k\rangle=\varepsilon_k^0|0\rangle+\varepsilon_k^1|1\rangle$ for $k=A,B$.
    Note that $\varepsilon$ is later related to the time step $\tau$.
    Importantly, in first order of $\varepsilon$, these tangent vectors can be approximated by product states
	\begin{equation}
    \begin{aligned}
		\label{eq:tangentialTPAprox}
        |\psi^{\prime}\rangle
        &\approx\left(|\psi_A\rangle+|\varepsilon_A\rangle\right)\otimes\left(|\psi_B\rangle+|\varepsilon_B\rangle\right).
    \end{aligned}
	\end{equation}

    Next, we relate these formal considerations to the open system evolution in Eq. \eqref{eq:FormalSolution}.
 
    Equation \eqref{eq:Lindblad} is invariant under the inhomogeneous transformation
    \begin{equation}
        H\mapsto H+\tfrac{1}{2i}\sum_{m}\big(\lambda^{m\ast} L^{m}-\lambda^{m}L^{m\dag}\big),\quad L^{m}\mapsto \lambda^{m}\mathbbm 1+L^{m},
    \end{equation}
    with $\lambda^{m}\in\mathbb{C}$. 
    This allows us to write all Kraus operators
    as perturbation of the identity $K^m=\mu^m(\mathbbm{1}+\varepsilon F^m)$ for each $m$, where $\varepsilon$ is a small parameter (later to be identified with the infinitesimal time step $\tau$) and $F^0=G/\|G\|$, $F^m=L^m/\|L^{m}\|$ are unit-norm operators. 
    Here, we defined
    \begin{equation}
        G=\frac{1}{i}H -\frac{1}{2}\sum_{m}\big(|\lambda^{m}|^2\mathbbm 1+2\lambda^{m\ast}L^{m}+L^{m\dag} L^{m}\big),
    \end{equation}
    $\mu^m=\sqrt{\tau}\lambda^m$, $F^m=L^m/\|L^{m}\|$, and $\lambda^m=\|L^m\|/\varepsilon$.

    Introducing projectors $\Pi_k=\tfrac{\ket{\psi_k}\bra{\psi_k}}{\braket{\psi_k|\psi_k}}$ and $\Pi_k^{\bot}=\mathbbm{1}-\Pi_k$ for $k=A,B$, allows us to expand the evolved state as 
    \begin{equation}
        \begin{split}
            K^m\ket{\psi}
            =&\mu^m\big(\mathbbm{1}+\varepsilon[\Pi_A^{\bot}+\Pi_A]\otimes[\Pi_B^{\bot}+\Pi_B]F^m\big)\ket{\psi}\\
            =&\mu^m(1+\varepsilon\braket{F^m})\ket{\psi}+ \mu^m\varepsilon \Pi_A^{\bot}(F^m)_A\ket{\psi}\\
            &+\mu^m\varepsilon\Pi_B^{\bot} (F^m)_B\ket{\psi}+\mu^m\varepsilon\Pi_A^{\bot}\otimes \Pi^{\bot}_B F^m\ket{\psi},
        \end{split}
        \label{eq:expansion}
    \end{equation}
    using the partially reduced operators \cite{SV13}
    $\left(F^m\right)_A=\tfrac{\langle\psi_{B}|F^m
        |\psi_{B}\rangle}{\langle\psi_B|\psi_B\rangle}
    $ and $\left(F^m\right)_B=\tfrac{\langle\psi_{A}|F^m
        |\psi_{A}\rangle}{\langle\psi_A|\psi_A\rangle}
    $, and $\braket{F^{m}}=\langle\psi|F^m|\psi\rangle$.
    By comparing Eq. \eqref{eq:expansion} with Eq. \eqref{eq:tangentialTP}, one observes that the last term $\mu^m\varepsilon\Pi_A^{\bot}\otimes \Pi^{\bot}_B F^m\ket{\psi}$ is not contained in the tangent space, thus not part of $\ket{\psi^\prime}=\ket{\psi^m(t+\tau)}$, implying
    \begin{equation}
        \begin{split}
            \ket{\psi^m(t+\tau)}=&\mu^m(1+\varepsilon\braket{F^m})\ket{\psi}+ \mu^m\varepsilon \Pi_A^{\bot}(F^m)_A\ket{\psi}\\
            &+\mu^m\varepsilon\Pi_B^{\bot} (F^m)_B\ket{\psi}.
        \end{split}
    \end{equation}
    Thus, utilizing Eq. \eqref{eq:tangentialTPAprox}, one obtains a product state
    \begin{align}
        \nonumber
        \label{eq:approx-tangent}
        \ket{\psi^m(t+\tau)}
        &\approx\mu^m(1+\varepsilon\langle F^m \rangle)\left(\ket{\psi_A}+\frac{\varepsilon\Pi_A^\bot (F^m)_A}{1+\varepsilon\langle F^m\rangle}\ket{\psi_A}\right)\\
        \nonumber&\otimes\left(\ket{\psi_B}+\frac{\varepsilon\Pi_B^\bot (F^m)_B}{1+\varepsilon\langle F^m\rangle}\ket{\psi_B}\right),\\
        \nonumber&=\frac{\mu^m}{1+\varepsilon\braket{F^m}}\big(\mathbbm{1}+\varepsilon(F^m)_A\big)\otimes \big(\mathbbm{1}+\varepsilon(F^m)_B\big)\ket{\psi},
        \\
        &=\frac{(K^m)_A\otimes (K^m)_B}{\braket{K^{m}}}\ket{\psi},
    \end{align}
    which is exact to first order in $\varepsilon$.
    Notably, because $\tau$ in Eq. \eqref{eq:FormalSolution} can be chosen of the same order as $\varepsilon$, we can view $\tfrac{1}{\braket{K^{m}}}(K^m)_A\otimes (K^m)_B\ket{\psi}$ as an open-system evolution in which the state remains separable.
    This approach removes any entanglement from the full evolution [Eq. \eqref{eq:FormalSolution}], while preserving all classical correlations \cite{PM24}.
    The corresponding density operator reads
	\begin{equation}
        \label{eq:SepMap}
        \rho(t+\tau)=\sum_{m} \frac{\left(K^m\right)_A\otimes \left(K^m\right)_B}{\braket{K^{m}}}\rho(t)\frac{\left(K^m\right)_A^\dag\otimes \left(K^m\right)_B^\dag}{\braket{K^{m}}^*},
	\end{equation}
    which describes a separable map as sought.
    That is, it is based on the physical open-system process under study while preserving separability.
      
\paragraph*{Separability Lindblad equation.---}

    The separable dynamics in Eq. \eqref{eq:SepMap} is explicitly given through the reduced operators
    \begin{equation}
        \begin{split}
            (K^0)_A
            &{=}
            \mathbbm{1}{-}i\tau(H)_A{-}\tfrac{\tau}{2}\sum_m\big(L^{m\dag}L^m\big)_A,
            ~
            (K^m)_A
            {=}\sqrt{\tau}(L^m)_A,
            \\
            (K^0)_B
            &{=}
            \mathbbm{1}{-}i\tau(H)_B{-}\tfrac{\tau}{2}\sum_m\big(L^{m\dag}L^m\big)_B,
            ~
            (K^m)_B{=}\sqrt{\tau}(L^m)_B,
        \end{split}
    \end{equation}
    for the Hamiltonian $H$ and Lindblad operators $L^m$, respectively.
    Comparing the evolved state in Eq. \eqref{eq:SepMap} with $\rho(t+\tau)\approx \rho+\tau \frac{d \rho}{d t}$, while keeping only contributions up to first order in $\tau$, yields an expression for the derivative, viz.
    \begin{equation}
        \label{eq:SepLin}
        \begin{split}
        \frac{d \rho}{d t}&=i\left[\rho,\left(H\right)_A\otimes \mathbbm{1}+\mathbbm{1}\otimes \left(H\right)_B\right]+\sum_{m}\left\langle L^{m\dag} L^{m}\right\rangle\rho\\
        &-\frac{1}{2}\sum_{m}\big\{\big(L^{m\dag} L^{m}\big)_A\otimes \mathbbm{1}+\mathbbm{1}\otimes\big(L^{m\dag} L^{m}\big)_B,\rho\big\}\\
         &+\sum_{m} \frac{\left(L^m\right)_A\otimes \left(L^m\right)_B}{\braket{L^{m}}}\rho\frac{\left(L^m\right)_A^\dag\otimes \left(L^m\right)_B^\dag}{\braket{L^{m}}^*}=\mathcal{L}_\rho(\rho).\\
        \end{split}
    \end{equation}
    We refer to this equation of motion as the \textit{separability Lindblad equation}.
    It describes the evolution of a composite system being restricted to a separable trajectory. 
    The solution of these equations can be compared with those of the conventional Lindblad equation to study the effect of entanglement in the dynamics of composite open quantum systems. 

\paragraph*{Properties and discussion.---}

    Firstly, Eq. \eqref{eq:SepLin} is nonlinear since both the partially reduced operators $(L^m)_k$ and $(H)_k$, for $k=A,B$, as well as the mean values in Eq. \eqref{eq:SepLin} depend explicitly on the state $\rho$.
    Secondly, Eq. \eqref{eq:SepLin} is a non-deterministic equation, (see the below discussion via stochastic differential equations) as the generator $\mathcal{L}_\rho$ is only defined for a pure product state $\rho=\ket{\psi_A\psi_B}\bra{\psi_A\psi_B}$.
    In general, dissipation leads $\rho$ to evolve into an ensemble of pure states $|\psi^m\rangle$ after an infinitesimal time step $\tau$, which is given by $\rho(t+\tau)=\rho+\tau \mathcal{L}_\rho(\rho)$.
    Each pure state in the ensemble can then be propagated by similar iterative steps, according to Eq. \eqref{eq:SepLin}.
    Thus, the evolved state is a mixture of all pure-state trajectories.
    In this sense, our equation describes a piece-wise deterministic process \cite{BP02} in which the generator $\mathcal{L}_\rho$ is updated with each iteration.
    In many practical calculations, we implement this via a Monte Carlo wave-function approach \cite{DZR92,MCD93,PK98} using the reduced operators in Eq. \eqref{eq:SepMap}. 
    Interestingly, Eq. \eqref{eq:SepLin} turns into a deterministic equation
    if the evolution is unitary \cite{SW17}, i.e., $L^m=0$ for all $m$, or if $\rho(t)$ is a decoherence-free state \cite{DG97,KBL01}.
    Moreover, the Lindblad operators are generally non-unique, i.e., the conventional master equation \eqref{eq:Lindblad} is invariant under the transformation $L^m\mapsto \sum_j u_{jm} L^j$, where $u_{jm}$ are the components of a unitary matrix. 
    We stress that the separability Lindblad equation \eqref{eq:SepLin} certifies entanglement regardless of the unraveling used for the Lindblad operators because the restricted state in Eq. \eqref{eq:SepMap} differs from the physical time evolution only in order $\varepsilon^2$, which is negligible; see Eq. \eqref{eq:approx-tangent}.

    When the dynamical map under study is separable, the separability-restricted dynamics [Eq. \eqref{eq:SepLin}] is identical to the unconstrained dynamics [Eq. \eqref{eq:Lindblad}], thus preserving all classical correlations. 
    The validity of this statement is evident by recognizing that for a separable map $\Lambda_t$, the non-Hermitian Hamiltonian $\hat{H}$ has to be local, i.e., $\hat{H}=\hat{H}_A\otimes\mathbbm{1}+\mathbbm{1}\otimes \hat{H}_B$, and the Lindblad operators must be in product form, i.e., $L^m=L^m_A\otimes L^m_B$.
    Then, the separability Lindblad equation \eqref{eq:SepLin} coincides with the regular Lindblad equation \eqref{eq:Lindblad}; see Ref. \cite{PAH24} for details.
    More generally, given a process $\Lambda_t$ and an initial product state $\rho_0$ for which $\Lambda_t(\rho_0)$ is separable at all times $t$, the dynamics coincides with the corresponding separability Lindblad equation \eqref{eq:SepLin}.
    Given a different initial product state $\tilde{\rho}_0$, the evolved state $\Lambda_t(\tilde{\rho}_0)$ might become entangled at some time $t$ and in this case Eq. \eqref{eq:SepLin} yields deviating quantum trajectories, signaling the build-up of entanglement.
    For example, choosing a single Lindblad operator as the controlled-NOT operator does not produce entanglement from the initial state $\ket{00}$ but creates entanglement when starting from $\tfrac{1}{\sqrt{2}}(\ket{0} + \ket{1})\otimes\ket{0}$; see Ref. \cite{PAH24} for details.
    In summary, the dynamics governed by the separability Lindblad equation coincide with the solution $\rho(t)$ of the conventional master equation \eqref{eq:Lindblad} if and only if $\rho(t)$ is separable at all times.
    Thus, our method provides a necessary and sufficient condition for entanglement in the evolution of an open quantum system.
    Also, witnessing entanglement in the dynamics would require to construct specific witnesses for each point in time, which is a rather challenging problem and not needed when applying our framework.

\paragraph*{Formulation as a stochastic equation.---}

    The separability Lindblad equation \eqref{eq:SepLin} corresponds to a piece-wise deterministic process \cite{BP02}, which can also be expressed as a stochastic differential equation.
    Specifically, we can introduce such an equation for the separable evolution \eqref{eq:SepMap} of a pure state $\sigma=\ket{\psi_A\psi_B}\bra{\psi_A\psi_B}$ as follows:
    \begin{equation}
        \label{eq:SepSto}
        \begin{split}
            d \sigma{=}&\bigg(i\left[\sigma,\left(H\right)_A{\otimes} \mathbbm{1}{+}\mathbbm{1}{\otimes} \left(H\right)_B\right]
            {+}2\sum_{m}\left\langle L^{m\dag}L^{m}\right\rangle\sigma
        \\
        &
            {-}\frac{1}{2}\sum_{m}\left\{\big(L^{m\dag} L^{m}\big)_A{\otimes} \mathbbm{1}{+}\mathbbm{1}{\otimes}\big(L^{m\dag} L^{m}\big)_B,\sigma\right\}\bigg)d t
        \\
        &
        {+}\sum_{m}\Bigg(
            \frac{\left(L^m\right)_A{\!\otimes} \left(L^m\right)_B\sigma\left(L^m\right)_A^\dag{\!\otimes} \left(L^m\right)_B^\dag}{\braket{L^{m}}\left\langle L^{m\dag} L^{m}\right\rangle\braket{L^{m}}^\ast}
            {-}\sigma
        \Bigg)d N_m,
        \end{split}
    \end{equation}
    where $\mathrm{d}N_m(t)=1$ and $d t=0$ at times where the quantum jump $m$ occurs; otherwise, we have $\mathrm{d}N_m(t)=0$.
    Moreover, the Poisson increments satisfy $d N_m(t)d N_n(t)=\delta_{mn}d N_n(t),$ and $\mathbb{E}[d N_m(t)]=\braket{L^{m\dag}L^m}d t$.
    Equation \eqref{eq:SepSto} is solved via the iteration $\sigma(t+dt)=\sigma(t)+d \sigma(t)$.
    As Eq. \eqref{eq:SepSto} preserves the purity of the state $\sigma$, the result of an open-system evolution is obtained by averaging over many trajectories $\sigma$, i.e., $\rho=\mathbb{E}[\sigma]$.
    Doing so, while making use of $d\rho=\mathbb{E}[d\sigma]$ and $\mathbb{E}[d N_m(t)]=\braket{L^{m\dag}L^m}d t$, one recovers the separability Lindblad equation \eqref{eq:SepLin}.
    If, additionally, the process under study is separable, Eq. \eqref{eq:SepSto} coincides with the common stochastic equation \cite{BP02,WM93}
    \begin{equation}
        \label{eq:stoch}
        \begin{split}
            d\sigma=&i\left(\sigma\hat{H}^\dag-\hat{H}\sigma\right)d t+\sum_m\braket{L^{m\dag}L^m}\sigma d t\\
            &+\sum_m\left(\frac{L^m\sigma L^{m\dag}}{\braket{L^{m\dag}L^m}}-\sigma\right)\mathrm{d}N_m.
        \end{split}
    \end{equation}

\begin{figure*}
    \includegraphics[width=\textwidth]{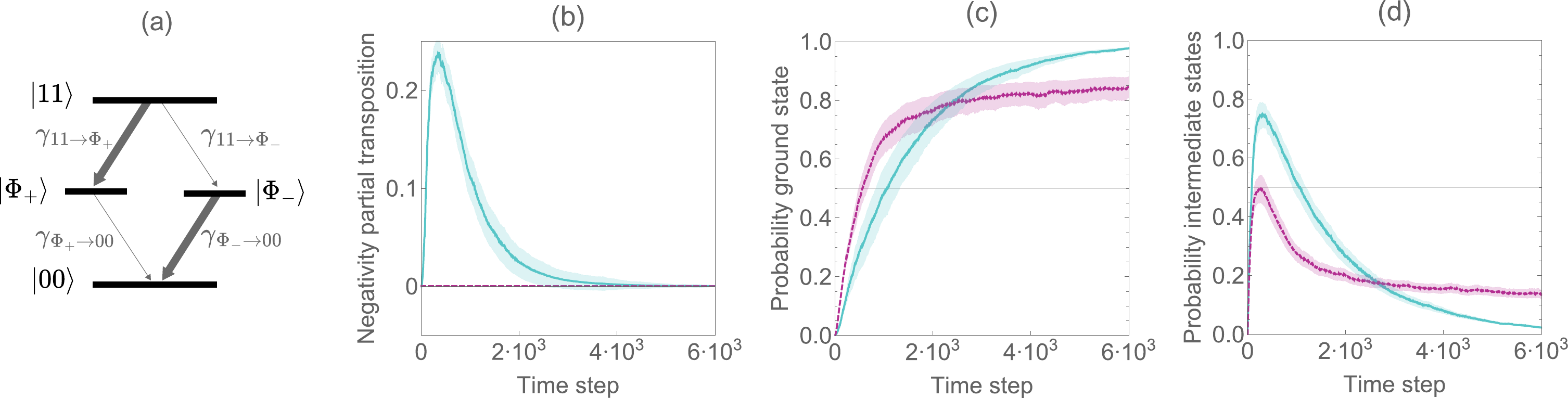}
	\caption{%
        \textbf{(a)} Scheme of the decay process given in Eq. \eqref{eq:exampleGenerators}.
        Comparison between the separable (dotted, purple line) and unrestricted (solid, turquoise line) solutions via Monte Carlo wave function.
		\textbf{(b)}
		Entanglement, characterized by the negativity of the partially transposed state.
		\textbf{(c)}
		Population of the ground state $\ket{00}$.
		\textbf{(d)}
		Probability to occupy one of the intermediate levels $|\Phi_+\rangle$ and $|\Phi_-\rangle$.
        One standard deviation uncertainty in lighter-colored bands.
        (Note that that the occupation of the $|11\rangle$ level can be inferred from (c) and (d).)
		Parameters for the numerical simulation are
		$\gamma_{11\to\Phi_+}=9=\gamma_{\Phi_-\to00}, \gamma_{11\to\Phi_-}=1=\gamma_{\Phi_+\to00}$, step size 
		$\tau=0.2$,
		and a sample size of $600$ \cite{Zenodo}. 
	}\label{fig:exampleLoss}
\end{figure*}

\paragraph*{Application: decay via Bell states.---}

	To focus on quantum-correlated dissipation, we consider a process with $H=0$ in which the initially doubly excited two-qubit state $|11\rangle$ decays into the ground state $|00\rangle$,
	and the process is mediated by decay channels with metastable levels in the form of Bell states $|\Phi_\pm\rangle=\tfrac{1}{\sqrt{2}}(|01\rangle\pm|10\rangle)$; see Fig. \ref{fig:exampleLoss} (a).
	The corresponding Lindblad operators are
	\begin{equation}
		\label{eq:exampleGenerators}
	\begin{aligned}
		&L^{1}=\sqrt{\gamma_{11\to\Phi_+}}|\Phi_+\rangle\langle 11|,\quad L^{2}=\sqrt{\gamma_{\Phi_+\to00}}|00\rangle\langle\Phi_+|,\\
        &L^{3}=\sqrt{\gamma_{11\to\Phi_-}}|\Phi_-\rangle\langle 11|,\quad L^{4}=\sqrt{\gamma_{\Phi_-\to00}}|00\rangle\langle\Phi_-|.
	\end{aligned}
	\end{equation}
	The transition rates $\gamma_{11\to\Phi_+}>\gamma_{11\to\Phi_-}$ are chosen so that most states decay into the entangled state $|\Phi_+\rangle$.
	In addition, $\gamma_{\Phi_+\to00}<\gamma_{\Phi_-\to00}$, rapidly depleting the few states in $|\Phi_-\rangle$. 
    Such situations are relevant to quantum-state engineering by dissipation \cite{VWC09} and more specifically, the four-level system in Fig. \ref{fig:exampleLoss} (a) can be realized in optical cavities \cite{KRS11}, plasmonic waveguides \cite{TCM11}, and using trapped ions \cite{CEZ22}.
    Beyond specific implementations, interactions of subsystems with a common heat bath \cite{B02(a)} can be used to harness environment induced entanglement \cite{BFP03,AMD10,VS10}.
    The separability Lindblad equation opens up an elegant route to study how successful the state engineering was in building up entanglement.

    To compare the restricted evolution \eqref{eq:SepLin} and the full evolution \eqref{eq:Lindblad}, we solve the respective Lindblad equations numerically via a Monte Carlo wave-function approach \cite{DZR92,MCD93,PK98}, using the Lindblad operators in Eq. \eqref{eq:exampleGenerators} and the partially reduced Lindblad operators $(L^m)_A,(L^m)_B$, respectively.
    The results are shown in Fig. \ref{fig:exampleLoss}.
    In panel (b), we witness entanglement building up over the evolution while the state remains separable in the restricted evolution; see also, e.g., Ref. \cite{BFP03}.
    To quantify the amount of entanglement, we utilize the negativity of the partially transposed state \cite{VW02}, i.e., the magnitude of its smallest negative eigenvalue.
    In both scenarios, there is a decay into the separable ground state $\ket{00}$, even though at different rates; see panel (c).
    The initial growth of entanglement coincides with the initially higher probability of the state $|\Phi_+\rangle$ being occupied; see graph (d).
    Moreover, the initial population and eventual depopulation are steeper for the entangling evolution than for its (restricted) separable counterpart.
    One can attribute this to a higher efficiency and speed of the process when not restricted to separable states \cite{YS22}.

\paragraph*{Application: random exchange interaction.---}

    The separability Lindblad equation does not only test whether a dynamical map is separable, but preserves all aspects of the dynamics which do not produce entanglement during the evolution.
    To see this, we consider a process with a single dissipator $L=\sqrt{\gamma} V$, where the swap operator $V\ket{\psi_A\psi_B}=\ket{\psi_B\psi_A}$ cannot be brought into product form.
    The unrestricted evolution is governed by
    \begin{equation}
        \label{eq:FullSwap}
        \frac{d \rho}{d t}=\gamma V\rho V -\gamma \rho.
    \end{equation}
    The corresponding dynamical map can be given analytically via exponentiation of the generator in Eq. \eqref{eq:FullSwap}, viz. 
    \begin{equation}
        \label{eq:FullSwapSol}
        \rho(t)=\me^{-\gamma t}\cosh(\gamma t)\rho+\me^{-\gamma t}\sinh(\gamma t)V\rho V.
    \end{equation}
    The evolution in Eq. \eqref{eq:FullSwapSol} corresponds to a random unitary map \cite{AS08}, in which with probability $\sinh(\gamma \tau){e}^{-\gamma \tau}$ the swap operation is applied.
    While the swap operation $V$ does not create entanglement, it is not separable;
    i.e., it cannot be brought into product form.
    Such maps are sometimes referred to as as non-entangling, but not completely non-entangling maps \cite{CV20}.
    Unable to introduce entanglement on their own, these maps have the capacity to generate multipartite entanglement when coupled to subsystems in which entanglement is already present \cite{VH05}.

    In order to obtain the restricted dynamics, we seek a solution of the separability Lindblad equation \eqref{eq:SepLin}.
    First, note that $(V)_A\ket{\psi_A}=\bra{\psi_B}V\ket{\psi_A\psi_B}=\braket{\psi_B\vert\psi_A} \ket{\psi_B}$,
    implies $(V)_A=\ket{\psi_B}\bra{\psi_B}$ for the reduced operator. 
    Normalization by $\braket{\psi_B|\psi_B}$ can be neglected.
    Likewise, we have $(V)_B=\ket{\psi_A}\bra{\psi_A}$. 
    Inserting these operators into Eq. \eqref{eq:SepLin} leads us to 
    \begin{equation}
        \label{eq:RedSwap}
        \begin{split}
            \frac{d \rho}{d t}&=\frac{\gamma}{\braket{V}^2}(V)_A\otimes (V)_B\rho (V)_A\otimes (V)_B-\gamma \rho,\\
        \end{split}
    \end{equation}
    where $\braket{V}=|\braket{\psi_A|\psi_B}|^2$.
    Equation \eqref{eq:RedSwap} has to be solved iteratively starting from $\ket{\psi_A\psi_B}\bra{\psi_A\psi_B}$. 
    After the first time step, the state propagates to
    $\rho(\tau)=(1-\gamma\tau)\ket{\psi_A\psi_B}\bra{\psi_A\psi_B}+ \gamma\tau\ket{\psi_B\psi_A}\bra{\psi_B\psi_A}$.
    The procedure continues for states at later time steps $s\tau$, with $s=2,3,\dots$
    Remarkably, the separability Lindblad equation can be solved analytically, viz.
    \begin{equation}
        \label{eq:SolSwap}
        \begin{split}
            \rho(s\tau)&=\sum_{a\text{ even}}^{s}{s\choose a}(1-\gamma \tau)^{s-a}(\gamma\tau)^a\ket{\psi_A\psi_B}\bra{\psi_A\psi_B}\\
            &+\sum_{a\text{ odd}}^{s}{s\choose a}(1-\gamma \tau)^{s-a}(\gamma\tau)^a\ket{\psi_B\psi_A}\bra{\psi_B\psi_A}.\\
        \end{split}
    \end{equation}
    Replacing $\tau=t/s$ and taking the limit $s\to\infty$ of infinitely many iteration steps, the restricted evolution \eqref{eq:SolSwap} coincides with the full evolution \eqref{eq:FullSwapSol}; see also Ref. \cite{PAH24}. 
    Thus, we have shown that our approach does not only verify whether a dynamical map is separable, but captures arbitrary non-entangling dynamics. 

\paragraph*{Conclusions.---}

    We devised a novel type of nonlinear Lindblad master equation that unambiguously identifies dynamical entanglement in open quantum systems.
    This was achieved by a projection approach, imposing the dynamics to yield separable trajectories only.
    When comparing these with the unconstrained trajectories, our method provides a necessary and sufficient condition for witnessing dynamical entanglement in open systems.
    For instance, we used this to show that entanglement is important for the effectiveness and speed in correlated decay mechanisms.
    Moreover, the separability Lindblad equation enables analytical treatment of dissipative processes as well, including the classification of dynamical quantum channels.

    Unlike existing approaches, entangling capabilities of a process are not characterized by input-output relations, but separability is imposed even at all intermediate times.
    The separability Lindblad equation further led to an equivalent stochastic equation, which is suitable for implementing numerical simulations.
    The separability Lindblad equation captures the most general type of non-entangling dynamics.
    It is therefore superior to analyzing separability of the Choi matrix of a dynamical map, which only tells us whether the channel has Kraus operators in product form.
    We further investigate these aspects in Ref. \cite{PAH24}.
    There, we also extend the analysis to multipartite systems of qudits, giving rise to a much richer structure and higher complexity than the bipartite qubit case.

    In summary, the here-derived open-system equations of motion cover all aspects of non-entangling dynamics that can be described by a Lindblad master equation.
    Our contribution is instrumental for studying fundamental properties of the dynamics regarding their classical and quantum capabilities in realistic settings and leads to a deeper understanding of time-dependent correlations in open quantum systems.

\paragraph*{Acknowledgments.---}

    The authors are grateful to Joan Alba for valuable suggestions and comments.
    The authors acknowledge funding through the Ministry of Culture and Science of the State of North Rhine-Westphalia (PhoQC initiative), the Deutsche Forschungsgemeinschaft (DFG, German Research Foundation) via the transregional collaborative research centers TRR 142 (Project C10, Grant No. 231447078), and the QuantERA project QuCABOoSE.
    J.P. acknowledges financial support from the Alexander von Humboldt Foundation (Feodor Lynen Research Fellowship).

\end{document}